\documentclass[10pt,twocolumn,twoside]{IEEEtran}

\usepackage[active]{srcltx} 
\usepackage[cmex10]{amsmath}
\usepackage{amsfonts}
\usepackage[final]{graphics}
\usepackage[final]{graphicx}
\usepackage{amsbsy}
\usepackage{amsbsy}
\usepackage{amssymb}
\usepackage{url}
\usepackage{enumerate}
\usepackage{cite}
\usepackage{psfrag}
\usepackage{color}
\usepackage{euscript}
\usepackage[utf8]{inputenc}
\usepackage{algorithmic}
\usepackage[ruled,vlined]{algorithm2e}

\newcommand{\bm}[1]{{\mathbf{#1}}}
\newcommand{\Es}{{\mathbb{E}}}          

\newcommand{\diag}{{\text{diag}}}

\newcommand{\I}{\bm{I}}

\newcommand{\ba}{\bm a}
\newcommand{\bg}{\bm g}

\newcommand{\fb}{\bm f}

\newcommand{\gb}{\bm g}

\newcommand{\Gammab}{\bm \Gamma}

\newcommand{\betab}{\boldsymbol \beta}
\newcommand{\gammab}{\boldsymbol \gamma}

\newcommand{\Bb}{\bm B}

\newcommand{\Cset}{\mathbb{C}}

\newcommand{\eqdef}{\triangleq}

\newcommand{\herm}{\text{H}}
\newcommand{\trasp}{\text{T}}

\newcommand{\pot}{\EuScript{P}}

\newcommand{\capaa}{\EuScript{C}}

\def\bdm#1\edm{\begin{displaymath}#1\end{displaymath}}
\def\be#1\ee{\begin{equation}#1\end{equation}}
\def\barr#1\earr{\begin{align}#1\end{align}}

\newcommand{\IeeeTCOMM}{{\em IEEE Trans.\ Commun.\/}}

\newcommand{\IeeeTWC}{{\em IEEE Trans.\ Wireless Commun.\/}}
\newcommand{\IeeeJSAC}{{\em IEEE J.\ Select.\ Areas Commun.\/}}

\newtheorem{theorem}{Theorem}[section]

\newtheorem{lemma}[theorem]{Lemma}

\begin{document}

\title{On the Capacity of Opportunistic Time-Sharing Downlink with 
a Reconfigurable Intelligent Surface}

\author{Donatella~Darsena,~\IEEEmembership{Senior Member,~IEEE}
and Francesco~Verde,~\IEEEmembership{Senior Member,~IEEE}
\thanks{
Manuscript received August 26, 2023; accepted October 4, 2023. Date
of publication xx yy  2023; date of current version xx yy 2023.
This work was partially supported by the European Union under the Italian National Recovery and Resilience Plan (NRRP) of NextGenerationEU, partnership on ``Telecommunications of the Future" (PE00000001 - program ``RESTART").
The associate editor coordinating the review of this article and approving it for
publication was Dr.~Gang Yang. ({\em Corresponding author: Francesco Verde.})
}
\thanks{
F.~Verde and D.~Darsena are with the Department of Electrical Engineering and Information Technology,  University Federico II, Naples I-80125,
Italy [e-mail: (f.verde, darsena)@unina.it].
The Authors are also with
National Inter-University Consortium for Telecommunications (CNIT).}
\thanks{Digital Object Identifier xxxxxxxxxxxxxxxxxxx.}
}
\markboth{IEEE Communications Letters,~Vol.~xx,
No.~yy,~Month~2023}{Darsena \MakeLowercase{\textit{et al.}}:
On the capacity of opportunistic time-sharing downlink with a reconfigurable intelligent surface}

\IEEEpubid{0000--0000/00\$00.00~\copyright~2023 IEEE}

\maketitle

\begin{abstract}

We provide accurate approximations of the sum-rate capacity 
of an opportunistic time-sharing downlink, when 
a reconfigurable intelligent surface (RIS) 
assists the transmission from a single-antenna base station (BS) 
to single-antenna user equipments (UEs).  
We consider the fading effects of both the direct 
(i.e., BS-to-UEs) and reflection (i.e, BS-to-RIS-to-UEs) links, 
by developing two approximations:
the former one is based on hardening of the reflection channel
for large values of the number of meta-atoms;
the latter one relies on  
the distribution of the sum of Nakagami variates
and does not require channel hardening.
Our derivations show
the dependence of the sum-rate capacity as a function of both
the number of users and the number of meta-atoms, as well as to establish a
comparison with a downlink without an RIS. 
Numerical results 
corroborate the accuracy and validity
of the mathematical analysis.

\end{abstract}


\section{Introduction}
\label{sec:intro}

\IEEEPARstart{I}{n} this letter, we consider a downlink channel 
in which a {\em reconfigurable intelligent surface (RIS)} is
employed to assist the transmission from a
single-antenna transmitter towards $K \gg 1$
single-antenna user equipments  (UEs).
An RIS is a metasurface composed of 
sub-wavelength meta-atoms, whose
reflection coefficients can be designed 
via software in order to suitably
manipulate the impinging signal \cite{Cui.2014}.
Relying on the feasibility of engineering 
the meta-atoms,  the wireless propagation 
environment might be programmed 
by optimizing on-the-fly the reflecting properties of 
an RIS to achieve different network-wide aims
\cite{DiRenzo.2020}.

When the base station (BS) can track 
the composite channels of the UEs,  
{\em opportunistic time sharing} is a simple and effective
transmission technique, which allows the BS to use time-division 
multiplexing and transmit to the best user. Such a scheduling strategy 
has been shown to achieve the
sum-rate capacity (maximum throughput) of the single-antenna
downlink channel \cite{Tse-book}.
A relevant question
to ask is the following: {\em How large of a performance boost does RIS-aided 
opportunistic time-sharing downlink provide over its conventional (i.e, without RIS) counterpart  in terms of sum-rate?}
The scaling law of the sum-rate
capacity of a Gaussian downlink with many users $K$ using opportunistic time sharing
has been deeply studied without an RIS \cite{Sharif.2007}. 
A similar study for an RIS-aided downlink with a large number of
users $K$ and meta-atoms $Q$ has not been carried out yet.

We focus on the sum-rate capacity
achievable using opportunistic time sharing
in RIS-aided downlinks. 
We develop an approximation of the sum-rate
capacity by invoking hardening 
of the reflection channel in the large $Q$ limit, which  
allows to readily unveil the scaling laws
as a function of $K$ and $Q$. 
Furthermore, we provide a very
accurate approximation of the sum-rate capacity
without invoking channel hardening, which is based 
on the sum of Nakagami variates. We also investigate
the interplay between the gain offered by the RIS and 
the selection diversity among the UEs.

\IEEEpubidadjcol

\section{Signal model and preliminaries}
\label{signal}

As in \cite{Wu_2019}, 
the downlink transmission among the BS and 
$K$ UEs is assisted by a digitally programmable 
RIS working in reflection mode, which is
made of  $Q=Q_x \times Q_y$ meta-atoms that 
can be independently and dynamically controlled by
digital logic devices \cite{Cui.2014}.
The meta-atoms are positioned along a rectangular grid 
having $Q_x$ and $Q_y$ elements on the $x$ and $y$ axes, respectively, with
constant inter-element spacing $d_\text{RIS}$. 
The channel between the BS and the RIS is assumed to
be characterized by a dominant line-of-sight component,
which is modeled as  
$\gb = \sigma_g \, \ba_\text{RIS}$,  
with pathloss $\sigma_g^2$
and signature
\begin{multline}
\ba_\text{RIS} \eqdef 
\left[1, e^{j \frac{2 \pi}{\lambda_0} d_\text{RIS} u_x}, \ldots, e^{j \frac{2 \pi}{\lambda_0} (Q_x-1) d_\text{RIS} u_x}\right]^\trasp
\\ \otimes  \left[1, e^{j  \frac{2 \pi}{\lambda_0} d_\text{RIS} u_y}, \ldots, e^{j  \frac{2 \pi}{\lambda_0} (Q_y-1) d_\text{RIS} u_y}\right]^\trasp \in \mathbb{C}^{Q}
\nonumber
\end{multline}
where   $\lambda_0=c/f_0$ is the wavelength, 
$c$ is the speed of the light in the medium, 
$\theta_\text{RIS} \in [0, 2 \pi)$ and 
$\phi_\text{RIS} \in [-\pi/2, \pi/2)$
identify the azimuth and elevation angles, respectively,
the corresponding directional cosines are 
$u_x \eqdef \sin \theta_\text{RIS} \, \cos \phi_\text{RIS}$
and $u_y \eqdef \sin \theta_\text{RIS} \, \sin \phi_\text{RIS}$, 
and $\otimes$ is the Kronecker product.

All the other relevant links are modeled as 
narrowband frequency-flat channels. Specifically,
for $k \in \{1,2,\ldots, K\}$,  $h_k \sim {\cal CN}(0, \sigma_{h_k}^2)$
models the low-pass equivalent channel 
response from the BS to UE $k$, whereas
$\fb_k \sim {\cal CN}(\bm{0}_Q, \sigma_{f_k}^2\, \I_{Q})$
represents  the low-pass equivalent channel 
response from the RIS to the $k$-th UE.
The parameters 
$\sigma_{h_k}^2$ and $\sigma_{f_k}^2$
are the large-scale geometric path losses of 
the links seen by the $k$-th UE 
with respect to the BS and the RIS, respectively.

We customarily assume that each UE uses standard timing synchronization with respect
to its direct link. After matched filtering and sampling at the baud rate,  
the discrete-time baseband  signal received at the $k$-th user reads as
\be
r_k   =  
c_{k}^* \left( \sum_{u=1}^K  \, \sqrt{\pot_u}\, s_{u} \right) +  v_{k}
\nonumber
\ee
where the {\em overall} channel gain seen by the $k$-th UE is  
$c_{k} \eqdef h_{k} + \gb^\herm \, \Gammab^* 
\, \fb_{k} \in \Cset$, for $k \in \{1,2,\ldots, K\}$,
with the $q$-th diagonal entry of 
$\Gammab \eqdef \diag\left(\gamma_1,
\gamma_2, \ldots, \gamma_Q \right)$ 
representing the reflection
coefficient of the $q$-th meta-atom,
$s_u$ being the information-bearing symbol intended for the
$u$-th user with corresponding transmit power 
$\pot_u$,  and $v_{k} \sim {\cal CN}(0,1)$ is the noise sample
at the output of the matched filter, with 
$v_{k_1}$ statistically
independent of $v_{k_2}$, for 
$k_1 \neq k_2$.
The transmitted symbols $s_1, s_2, \ldots, s_K$ are 
independent and identically
distributed (i.i.d.) complex circular zero-mean unit-variance random variables (RVs).
The couple $(h_k, \fb_k)$ is independent of both $v_k$
and $s_k$, $\forall k \in \{1,2,\ldots, K\}$. 

The  {\em sum-rate capacity} (in bits/s/Hz) is defined as follows
\be
\capaa_\text{sum} =  \max_{
\shortstack{\footnotesize $\pot_1, \pot_2, \ldots, \pot_K$ 
\\ \footnotesize $\gamma_1, \gamma_2, \ldots,\gamma_Q$}}  
\sum_{k=1}^K \log_2 \left( 1+ \frac{\pot_k \, |c_k|^2}{ 
|c_k|^2 \sum_{u \neq k} 
\pot_u +1}\right)
\nonumber
\ee
subject to the {\em transmit power constraint}
 $\sum_{k=1}^K  \pot_k \le \pot_\text{TX}$,
with  $\pot_\text{TX} >0$ being the (fixed) 
maximum allowed transmit power,
and the {\em global passivity constraint} 
$\|\gammab\|^2 \le Q$  at the RIS \cite{DiRenzo_IEEE_Proc_2022},
with $\gammab \eqdef [\gamma_1, \gamma_2, \ldots, \gamma_Q]^\herm \in \Cset^{Q}$.
For a lossless RIS, the latter constraint yields
$\|\gammab\|^2=Q$.\footnote{Another option for a lossless RIS
consists of imposing 
the $Q$ {\em local passivity constraints} $|\gamma_q|=1$,
$\forall q \in \{1,2,\ldots, Q\}$ \cite{Wu_2019}. However, 
in this case, optimization of the reflection response 
is a non-convex NP-hard problem. 
The best known methods that develop a solution for this problem are iterative
and do not provide a closed-form solution.}
 
Given  the reflection vector $\gammab$, the sum-rate capacity is
equal to the largest single-user capacity in the system \cite{Tse-book}, that is, the resource
allocation policy is the opportunistic time-sharing strategy:
$\pot_k =\pot_\text{TX}$ if $k=k_\text{max}$, 
$\pot_k =0$ otherwise,
with  $k_\text{max} \eqdef \arg \max_{k \in \{1,2,\ldots, K\}} |c_k|^2$.
Recalling the expression of $c_k$, the corresponding
sum-rate capacity boils down to 
\be
\capaa_\text{sum}=  
\log_2\left(1+ \pot_\text{TX} \, \alpha_\text{opt} \right) 
\label{eq:C-sum}
\ee
with
\be
\alpha_\text{opt} \eqdef 
\max_{
\shortstack{\footnotesize  $k \in \{1,2,\ldots,K\}$
\\ \footnotesize $\gammab \in \Cset^Q$: $\|\gammab\|^2=Q$}}  \, 
|h_{k}|^2 +  2 \, \Re\left\{
\betab_k^\herm \, \gammab\right\} + \gammab^\herm \, \Bb_k \, \gammab 
\label{eq:max-prob-ckm-relax}
\ee
where  
$\betab_k \eqdef h_k \, \diag(\fb_k^*) \, \bg  \in \Cset^{Q}$
and we have defined the Hermitian matrix $\Bb_k \eqdef \diag(\fb_k^*) \, \bg \, \bg^\herm \, \diag(\fb_k) \in \Cset^{Q \times Q}$.
Closed-form solution of \eqref{eq:max-prob-ckm-relax} is provided
by Lemma~\ref{lemma:2}.

\begin{lemma}
\label{lemma:2}
Under the constraint $\|\gammab\|^2=Q$,  
the cost function in \eqref{eq:max-prob-ckm-relax} can be upper bounded 
for each $k$ as follows
\be
|h_{k}|^2 +  2 \, \Re\left\{
\betab_{k}^\herm \, \gammab\right\} + \gammab^\herm \, 
\Bb_{k} \, \gammab \\ \le
\left(|h_{k}| + \sqrt{Q} \, \| \diag(\fb_{k}^*) \, \bg\|\right)^2
\nonumber 
\ee
where the equality holds if and only if
\be
\gammab = \sqrt{Q} \, \frac{h_{k}}{|h_{k}|} \, 
\frac{\diag(\fb_{k}^*) \, \bg}{\| \diag(\fb_{k}^*) \, \bg\|} \: .
\label{eq:gammaopt}
\ee
\end{lemma}

{\emph Proof:} The proof comes from: (a)
$\Re\{x\} \le |x|$ $\forall x \in \Cset$;
(b) the Cauchy-Schwarz inequality
$\left|\betab_k^\herm \, \gammab\right| \le  \|\betab_k\| \, \|\gammab\|$;  
(c) the Rayleigh-Ritz theorem and observing that 
the maximum eigenvalue of the rank-one matrix $\Bb_k$ is 
$\| \diag(\fb_k^*) \, \bg\|^2$.

\noindent Physically, solution \eqref{eq:gammaopt}
implies that the RIS may introduce local power 
amplifications (i.e., $|\gamma_q|>1$)  
or local power losses (i.e., $|\gamma_q|<1$)
for some meta-atoms, 
while ensuring that the total reradiated power by the lossless RIS is equal to the 
total incident power (see \cite{DiRenzo_IEEE_Proc_2022} for implementation details).

\section{Theoretical performance analysis}
\label{sec:perf}

By applying  channel coding across
channel coherence  intervals (i.e., over an ``ergodic" 
interval of channel variation with time), the {\em average} 
sum-rate capacity \cite{Tse-book} 
is given by
\be
\overline{\capaa}_\text{sum} \eqdef \Es[\capaa_\text{sum}] = 
\int_{0}^{+\infty} 
\log_2\left(1+ \pot_\text{TX} \, \alpha \right) \, f_{\alpha_\text{opt}}(\alpha) 
\, {\rm d} \alpha 
\label{eq:Ave-C-sum}
\ee
where $f_{\alpha_\text{opt}}(\alpha)$ is 
the probability density function (pdf) of the RV
$\alpha_\text{opt}$, which, by virtue of Lemma~\ref{lemma:2}, can
be explicitly written as 
$\alpha_\text{opt} =  \max_{k \in \{1,2,\ldots,K\}} \,
\left(|h_k| + \sqrt{Q} \, \| \diag(\fb_k^*) \, \bg\|\right)^2$.
Under the opportunistic time-sharing strategy, there is just one user transmitting 
at any time and, thus, we can resort to  
the encoding and decoding procedures for the code designed for a 
point-to-point channel \cite{Tse-book}.

We consider the case 
in which the users approximately experience the same large-scale 
geometric path loss, i.e., the parameters $\sigma_{h_k}^2$ and $\sigma_{f_k}^2$
do not depend on $k$, i.e., $\sigma_{h_k}^2 \equiv \sigma_{h}^2$ and
$\sigma_{f_k}^2 \equiv \sigma_{f}^2$, $\forall k \in \{1,2,\ldots, K\}$,
which will be referred to as the case of {\em homogeneous}
 users.\footnote{From
a physical viewpoint, this happens when the users form a 
{\em cluster}, wherein the distances between the different UEs are negligible
with respect to the distance between the transmitter and the RIS.}
In this case, the RV $\alpha_\text{opt}$ is the
maximum of $K$ i.i.d. RVs 
and its pdf is computed as  
\be
f_{\alpha_\text{opt}}(\alpha) = K \,  f_{X_k}(\alpha)  
\, \left[F_{X_k}(\alpha)\right]^{K-1} 
\label{eq:fopt}
\ee
where $f_{X_k}(\alpha)$ and 
$F_{X_k}(\alpha)$ denote the pdf and  
the cumulative distribution function (cdf) of the RV 
$X_k \eqdef Z_k^2$, with $Z_k \eqdef Z_k^{(1)}+Z_k^{(2)}$, 
$Z_k^{(1)} \eqdef |h_k|$, and $Z_k^{(2)} \eqdef \sigma_g \, \sqrt{Q} \, \|\fb_k\|$.
The distributions of $Z_k^{(1)}$ and $Z_k^{(2)}$ are discussed in 
Appendix~\ref{app:app-1}.

Trying to work with \eqref{eq:fopt} for 
evaluating \eqref{eq:Ave-C-sum}
is  numerically difficult even for small values of $K$.
Hence, we apply extreme value theory \cite{Lea.1978}
to calculate the distribution of $\alpha_\text{opt}$
when $K$ is sufficiently large.
Relying on the limit laws for maxima \cite{Lea.1978},
provided that the cdf of $X_k$ is 
a von Mises function,\footnote{It fulfills \cite{Kaly.2012} 
$\lim_{\alpha \to + \infty} 
\left[ \frac{1-F_{X_k}(\alpha)}{f_{X_k}^2(\alpha)}\right]
\, \frac{{\rm d}}{{\rm d}\alpha} f_{X_k}(\alpha)=-1$.}
as $K \to \infty$, the RV
$\alpha_\text{opt}$ convergences  in distribution \cite{Gumbel.1958} 
to the Gumbel distribution
$\lim_{K \to \infty} F_{\alpha_\text{opt}}(\alpha) =
e^{ - e^{- \frac{\alpha-b_K}{a_K}}}$,
where $F_{\alpha_\text{opt}}(\alpha)$
is the cdf of  $\alpha_\text{opt}$, whereas
\be
b_K   \eqdef F_{X_k}^{-1}\left(1-\frac{1}{K}\right) 
\; \text{and} \;\;
a_K  \eqdef 
\frac{1}{K \, f_{X_k}(b_K)} \: .
\label{eq:constant}
\ee
Replacing $f_{\alpha_\text{opt}}(\alpha)$ with the 
Gumbel pdf, eq.~\eqref{eq:Ave-C-sum} reads as
\be
\overline{\capaa}_\text{sum} \asymp \frac{1}{a_K}
\int_{0}^{+\infty} 
\log_2\left(1+ \pot_\text{TX} \, \alpha \right) \,
e^{- \frac{\alpha-b_K}{a_K}} \, e^{- e^{- \frac{\alpha-b_K}{a_K}}}
\, {\rm d} \alpha 
\label{eq:Ave-C-sum-asympt}
\ee
with $x \asymp y$ indicating that 
$\lim_{K \to + \infty} x/y =1$.\footnote{By using the Maclaurin 
series of the exponential function,  the integral \eqref{eq:Ave-C-sum-asympt} can be 
rewritten as an absolutely convergent  series, which  
can be approximately evaluated by using a finite number 
of terms \cite[Appendix~C]{Kaly.2012}.}

The pdf or cdf of $X_k$ must be derived to evaluate 
$\overline{\capaa}_\text{sum}$ by using either 
\eqref{eq:Ave-C-sum}-\eqref{eq:fopt} or 
\eqref{eq:Ave-C-sum-asympt}.
To approximate the distribution of $X_k$, 
we distinguish two cases: in Subsection~\ref{sec:Case_1}, we assume that
the reflection channel from the RIS to each UE hardens for 
sufficiently large values of $Q$; in Subsection~\ref{sec:Case_2}, we derive
a more general approximation that does not require 
hardening of the reflection channel.  
In Subsection~\ref{sec:dis}, we derive the average
{\em receive} signal-to-noise ratio (SNR) of the user selected for scheduling,
which allows us to discuss the impact of the parameters $K$ and $Q$ on system performance.

\subsection{Approximation 1: Hardening of the reflection channel}
\label{sec:Case_1}

The $k$-th reflection channel hardens \cite{Ngo.2017} if
$Z_k^{(2)}/\Es[Z_k^{(2)}]$ converges in probability to $1$,
as $Q \to + \infty$, for each $k \in \{1,2,\ldots, K\}$.
Based on the Markov inequality \cite{Casella},  a sufficient condition 
for the hardening of $Z_k^{(2)}$ is $\text{VAR}[Z_k^{(2)}]/\Es^2[Z_k^{(2)}]
\to 0$, as $Q \to + \infty$. 
By virtue of \eqref{mean-Naka} and \eqref{var-Naka}, one gets
$\text{VAR}[Z_k^{(2)}]/\Es^2[Z_k^{(2)}] \approx 1/(4 \, Q)$ for 
very large $Q$. Therefore, as $Q \to + \infty$,  the pdf of $Z_k^{(2)}$
can be approximated by the following Dirac delta distribution
$f_{Z_k^{(2)}}(\alpha) \approx \delta\left(\alpha-\Es[Z_k^{(2)}]\right) $,
with the mean of $Z_k^{(2)}$ given by \eqref{mean-Naka}.
Accordingly, relying on the results of the transformations of RVs \cite{Proakis},
the pdf of $X_k$ is approximated by
\be
f_{X_k}(\alpha) \approx  
\frac{\sqrt{\alpha}-\Es[Z_k^{(2)}]}{\sigma_{h}^2 \, \sqrt{\alpha}}
\, e^{-\frac{\left(\sqrt{\alpha}-\Es[Z_k^{(2)}]\right)^2}{\sigma_{h}^2}} \:,
\quad \text{for $\alpha > \Es[Z_k^{(2)}]$} \:.
\nonumber
\ee
After algebraic manipulations, the cdf of  $X_k$ is 
cor\-re\-spond\-ing\-ly approximated:
\be
F_{X_k}(\alpha) \approx
1- e^{-\frac{\left(\sqrt{\alpha}-\Es[Z_k^{(2)}] \right)^2}{\sigma_{h}^2}} \:,
\quad \text{for $\alpha > \Es[Z_k^{(2)}]$} \:.
\label{eq:cdfXhat}
\ee

For any value of $K$, 
the above distributions can be substituted
in  \eqref{eq:fopt} in order to obtain an approximation of the pdf of 
$\alpha_\text{opt}$, which is involved in the calculus 
of the average sum-rate capacity \eqref{eq:Ave-C-sum}.
On the other hand, since the distribution \eqref{eq:cdfXhat} is a von Mises function, one can resort to 
\eqref{eq:Ave-C-sum-asympt} for sufficiently large values of $K$, where
\eqref{eq:constant} ends up to
\be
b_K  \approx \left[\sigma_{f} \, \sigma_{g} \, Q+ \sigma_{h} \sqrt{\ln(K)}\right]^2
\; \text{and} \;\; 
a_K  \approx \sigma_{h}^2 + \frac{\sigma_{f} \, \sigma_{g} \, \sigma_{h} \, Q}{\sqrt{\ln(K)}} \:.
\label{eq:constant-2}
\ee

\subsection{Approximation 2: Sum of Nakagami variates}
\label{sec:Case_2}

The RV $Z_k$ is the sum of the two independent
and non-identically distributed (i.n.i.d.) Nakagami RVs 
$Z_k^{(1)}$ and $Z_k^{(2)}$. Therefore, the pdf
of $Z_k$ is the convolution of the pdfs of $Z_k^{(1)}$ and $Z_k^{(2)}$,
which does not admit a closed-form expression.
Following \cite{Hadzi.2009}, we propose to approxi\-mate 
the pdf $f_{Z_k}(\alpha)$ of $Z_k$ by the pdf $f_{\widehat{Z}_k}(\alpha) $ 
of  the RV  
$\widehat{Z}_k \eqdef \sqrt{[\widehat{Z}_k^{(1)}]^2+[\widehat{Z}_k^{(2)}]^2}$,
where $\widehat{Z}_k^{(1)}$ and $\widehat{Z}_k^{(2)}$ are i.i.d. 
Nakagami RVs with shape parameter $\widehat{m}$
and scale parameter $\widehat{\Omega}$, which are determined such that 
$f_{\widehat{Z}_k}(\alpha) $ be an
accurate approximation of $f_{Z_k}(\alpha) $. 
The choice of $\widehat{m}$
and $\widehat{\Omega}$ is discussed in Appendix~\ref{app:app-2}.

At this point, we would like to point out that the main advantage 
of using $f_{\widehat{Z}_k}(\alpha) $
{\em in lieu} of  $f_{Z_k}(\alpha) $ stems from the fact that 
the square of  $\widehat{Z}_k$ is the sum of 
two i.i.d. gamma RVs \cite{Casella}. Indeed, the 
square of a Nakagami RV with 
shape parameter $\widehat{m}$
and scale parameter $\widehat{\Omega}$ turns out to be 
a gamma RV with shape parameter 
$\widehat{m}$ and scale parameter $\widehat{\Omega}/\widehat{m}$.
Moreover, the sum of the two i.i.d. gamma RVs 
$[\widehat{Z}_k^{(1)}]^2$ and $[\widehat{Z}_k^{(2)}]^2$
is a gamma RV, too, with 
shape parameter 
$2 \, \widehat{m}$ and scale parameter $\widehat{\Omega}/\widehat{m}$.
Thus, we can conclude that the pdf of the RV 
$\widehat{X}_k \eqdef \widehat{Z}_k^2$ is given by (see, e.g., \cite{Casella})
\be
f_{\widehat{X}_k}(\alpha)= \left(\frac{\widehat{m}}{\widehat{\Omega}}\right)^{2 \, \widehat{m}}
\, \frac{\alpha^{2 \, \widehat{m}-1}}{\Gamma(2 \, \widehat{m})} \, e^{- \frac{\widehat{m}}{\widehat{\Omega}} \alpha} \:, \quad \text{for $\alpha>0$} 
\label{eq:Gamma-pdf}
\ee
whose cdf reads as
$F_{\widehat{X}_k}(\alpha) = P \left( \frac{\widehat{m} \, \alpha}{\widehat{\Omega}}, 2 \, \widehat{m}\right)$,\footnote{It can be verified that $F_{\widehat{X}_k}(\alpha)$
is a von Mises function, too.} 
where 
$P(x,a) \eqdef  \frac{1}{\Gamma(a)} \int_{0}^{x} t^{a-1} \, e^{-t} \, {\rm d}t$
is the (regularized) lower incomplete gamma function
and $\Gamma(a)$ is defined in  Appendix~\ref{app:app-1}.

According to
\eqref{eq:constant}, the 
constant $b_K$ is the 
solution of 
$P \left( \frac{\widehat{m} \, b_K}{\widehat{\Omega}}, 2 \, \widehat{m}\right)=1-\frac{1}{K}$,
which is given by 
\be
 b_K \approx \frac{\widehat{\Omega}}{\widehat{m}} \, 
 P^{-1} \left(1-\frac{1}{K}, 2 \, \widehat{m}\right)
\label{eq:b_K-naka}
\ee
where $P^{-1}(y,a)$ is the inverse of the lower incomplete gamma function
for $y \in [0,1]$, i.e., $P^{-1}(P(x,a),a)=x$.
Substituting \eqref{eq:b_K-naka} in \eqref{eq:constant} and replacing  
$f_{X_k}(b_K)$ with $f_{\widehat{X}_k}(b_K)$, one has
\be
a_K \approx \frac{\widehat{\Omega}}{\widehat{m}} \, 
\frac{\Gamma(2 \, \widehat{m})}{K  \, 
\left[P^{-1} \left(1-\frac{1}{K}, 2 \, \widehat{m}\right)\right]^{2 \, \widehat{m}-1} \, e^{-P^{-1} \left(1-\frac{1}{K}, 2 \, \widehat{m}\right)}} \:.
\label{eq:a_K-naka}
\ee

\subsection{Average receive SNR}
\label{sec:dis}

The asymptotic Gumbel distribution 
allows one to statistically characterize  
the receive SNR of the user selected for scheduling, which 
is defined as
$\rho_\text{sum} \eqdef \pot_\text{TX} \, \alpha_\text{opt}$. 
The average receive
SNR $\overline{\rho}_\text{sum}\eqdef \Es[\rho_\text{sum}]
= \pot_\text{TX} \, \Es[\alpha_\text{opt}]$ can be derived 
from the mean of the Gumbel distribution, thus yielding 
\be
\overline{\rho}_\text{sum} =\pot_\text{TX} \, ( b_K + C \, a_K)
\label{eq:avesnrr}
\ee
with 
$C \approx 0.5772$ being the Euler-Mascheroni constant.

In a downlink {\em without} an RIS, the RV 
$X_k=[Z_k^{(1)}]^2=|h_k|^2$ is exponentially distributed with mean $\sigma_h^2$.
In this case, it follows from \eqref{eq:constant} that $b_K= \sigma_h^2 \, \ln (K)$ and 
$a_K= \sigma_h^2$. From \eqref{eq:avesnrr}, the average receive SNR 
in an RIS-unaided opportunistic time-sharing downlink is 
\[
\overline{\rho}_\text{sum}^{\,\text{w/o RIS}} =\pot_\text{TX} \, \sigma_h^2 
\, [C+\ln(K)]  \, \Rightarrow \lim_{K \to + \infty}  \frac{\overline{\rho}_\text{sum}^{\,\text{w/o RIS}}}{\ln(K)}=\pot_\text{TX} \, \sigma_h^2 
\]
which benefits by a factor of $\ln(K)$ asymptotically for large $K$
(so-called {\em multiuser diversity effect}), and
$\overline{\capaa}_\text{sum}^{\,\text{w/o RIS}}$ 
increases double logarithmically in $K$ \cite{Sharif.2007}. 

For a RIS-aided downlink, let us 
first consider the case when $K \to + \infty$ 
and the reflection channel hardens. In this situation, 
using  \eqref{eq:constant-2},
the average receive SNR in \eqref{eq:avesnrr} is approximated as
\[
\overline{\rho}_\text{sum} \approx \overline{\rho}_\text{sum}^{\,\text{w/o RIS}}   +
\pot_\text{TX} \left[ \sigma_f^2 \, \sigma_g^2 \, Q^2  
+ \sigma_f \, \sigma_g \, \sigma_h \,  \frac{2 \, \ln(K)+C}{\sqrt{\ln(K)}} \, Q\right]
\label{eq:avesnr-hardening} 
\]
which shows that, compared to the RIS-unaided downlink,  
the SNR gain provided by the presence of the RIS depends on 
the relationship between $K$ and $Q$.
Indeed, if  $Q$ approaches infinity at the same rate as $K$, i.e.,
$Q=\chi \, K$, where $\chi \neq 0$ is a constant independent
of $K$, it results that 
$\lim_{K,Q \to + \infty}
{\overline{\rho}_\text{sum}}/{Q^2}= \pot_\text{TX}  \, \sigma_f^2 \, \sigma_g^2$,
i.e.,  the average receive SNR scales like $Q^2$
as $K$ and $Q$ grow to infinity. In this case,
the downlink performance is dominated by the 
reflection channel and the 
reflection process of the RIS becomes predominant 
with respect to multiuser diversity effects.
On the other hand, if $Q$ approaches $+\infty$
as $Q = \chi \, \sqrt{\ln(K)}$, one has
$\lim_{K,Q \to + \infty}
{\overline{\rho}_\text{sum}}/{\ln(K)}= 
\pot_\text{TX}  \left[(\sigma_f \, \sigma_g \, \chi+\sigma_h)^2
+ C \, \sigma_f \, \sigma_g \, \sigma_h \, \chi \right]$.
In this case, the sum-rate capacity also depends on the
direct channel and the effect of the RIS
becomes negligible when 
$\sigma_f \, \sigma_g \, \chi \ll \sigma_h$, thus 
approaching the performance of the 
RIS-unaided downlink.
 
When 
$K$ grows to infinity and hardening 
of the reflection channel does not hold,
we resort to 
\eqref{eq:Gamma-pdf} for approximating the distribution of $X_k$.
In this case, the average receive SNR can be obtained
by substituting \eqref{eq:b_K-naka} and \eqref{eq:a_K-naka}
in \eqref{eq:avesnrr}, whose 
dependence on 
$K$ and $Q$ will be shown numerically in the forthcoming Section~\ref{sec:simulations}.

\begin{figure}[!t]
\centering
\includegraphics[width=0.95\linewidth, trim=10 20 40 20]{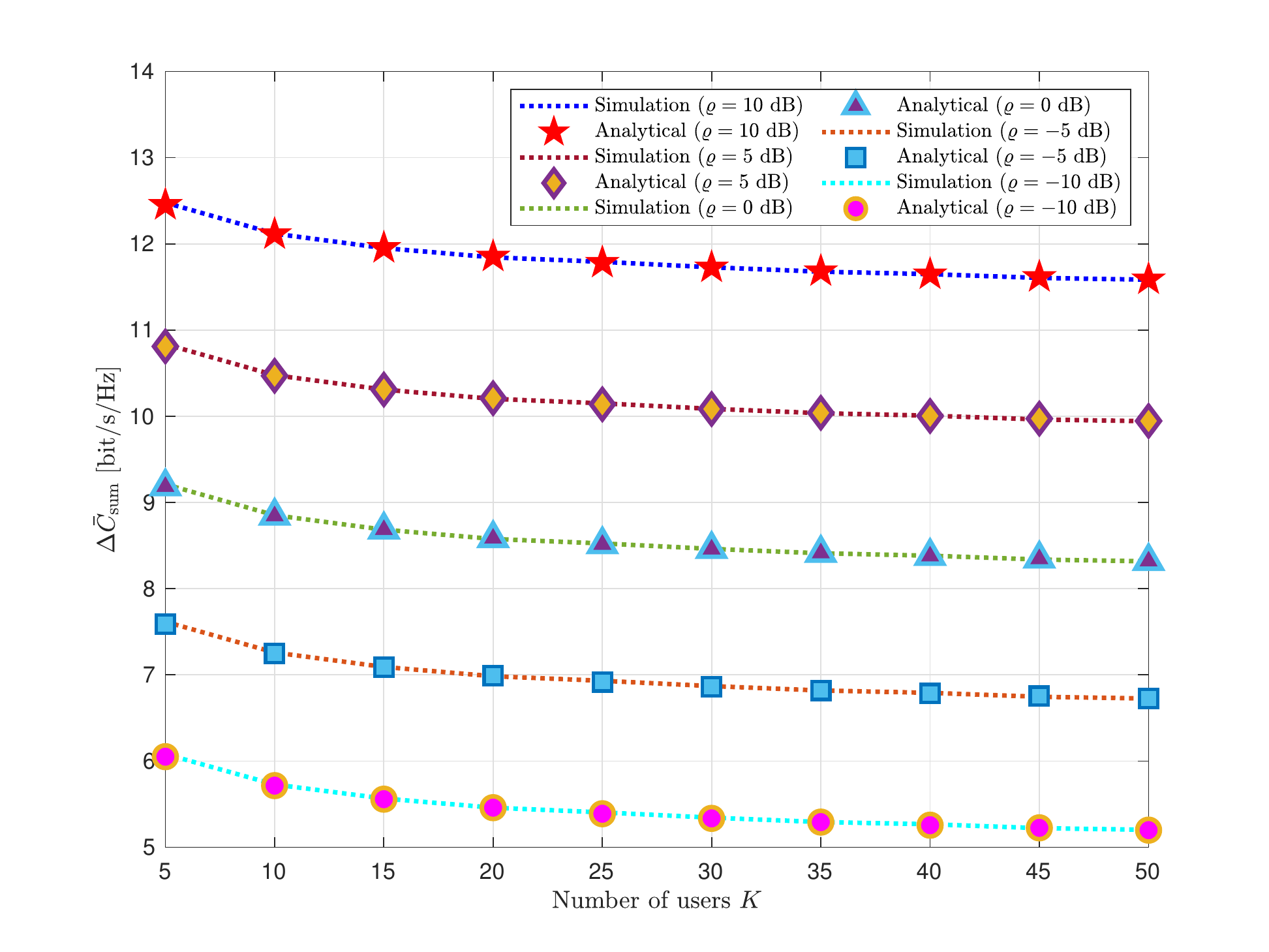} 
\caption{$\Delta \overline{\capaa}_\text{sum}$ versus $K$ (Approximation 2 for the analytical curves).}
\label{fig:fig_1}
\end{figure}

\section{Numerical performance analysis}
\label{sec:simulations}

We consider a $2$-D Cartesian system, wherein the BS and 
the RIS are located at $(0,0)$ and $(10,0)$ (in meters), respectively, 
whereas the users form a circular cluster centered in $(40, -10)$ (in meters).
The inter-element spacing is fixed to $d_{\mathrm{RIS}} = \lambda_0/4$, whereas the azimuth and elevation angles at the RIS are uniformly distributed in 
$[0, 2 \pi)$ and $[-\pi/2, \pi/2)$, respectively.
All the other channel links are independently generated by assuming a carrier frequency $f_0 = 25$ GHz, with 
variance $\sigma^2_{\alpha} = G_{\alpha} \, d_{\alpha}^{-\eta} \, \lambda_0^2/(4 \pi)^2$, for $\alpha \in \{g, h\}$, where $G_{\alpha} = 25$ dBi for the RIS and $G_{\alpha} = 5$ dBi for the UEs, while $d_{\alpha}$ represents the distance of the link and $\eta = 1.6$ is the path loss exponent. 
The variance $\sigma^2_f$ of the channel link between the RIS and the UEs is derived from the ratio $\varrho \eqdef (\sigma^2_f \, \sigma^2_g) /\sigma^2_h$, which assumes  the 
values in $\{0, \pm 5, \pm 10\}$ dB.
The effective isotropic radiated power of the BS is set to $33$ dBm and the noise power at the UEs is equal to $-100$ dBm. 

In Fig.~\ref{fig:fig_1}, we report the difference $\Delta \overline{\capaa}_\text{sum} \eqdef 
\overline{\capaa}_\text{sum}-\overline{\capaa}_\text{sum}^\text{w/o RIS}$
between the average sum-rate capacity of the RIS-aided  
and RIS-unaided  downlinks, as a function of the number of users $K$
for different values of $\varrho$.
The rates $\overline{\capaa}_\text{sum}$ and $\overline{\capaa}_\text{sum}^\text{w/o RIS}$
are numerically obtained by averaging \eqref{eq:C-sum} 
over $1000$ independent Monte Carlo runs, by setting  
$Q=30$ for the RIS-aided downlink and $Q=0$ for the RIS-unaided one,
respectively. The corresponding analytical curves 
\eqref{eq:Ave-C-sum-asympt}  are plotted by using
the parameters \eqref{eq:b_K-naka} and \eqref{eq:a_K-naka} (Approximation 2).
Besides corroborating the noticeable accuracy of the proposed approximation, which does not
require hardening of the reflection channel, it is seen from Fig.~\ref{fig:fig_1}
that $\Delta \overline{\capaa}_\text{sum}$ decreases with $K$. This behavior is due
to the fact that the increase in $K$ of $\overline{\capaa}_\text{sum}$ is
partially hidden by the larger SNR gain due to reflection process of the RIS, while 
$\overline{\capaa}_\text{sum}^\text{w/o RIS}$ scales like $\ln(\ln (K))$.

\begin{figure}[!t]
\centering
\includegraphics[width=0.95\linewidth, trim=10 20 40 20]{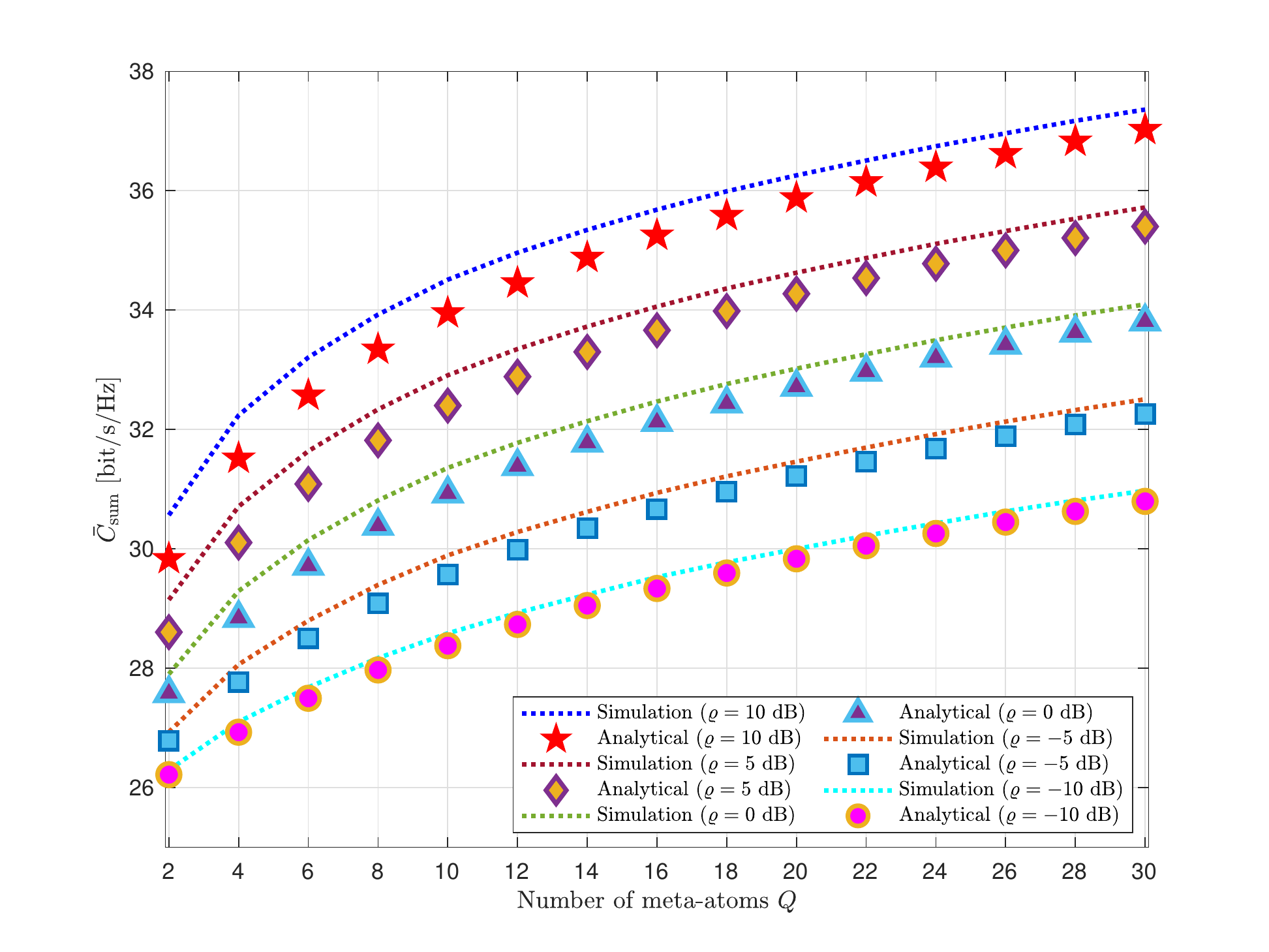}
\caption{$\overline{\capaa}_\text{sum}$ versus $Q$ ($K=10$, Approximation 1 for the analytical curves).}
\label{fig:fig_2}
\end{figure}

\begin{figure}[!t]
\centering
\includegraphics[width=0.95\linewidth, trim=10 20 40 20]{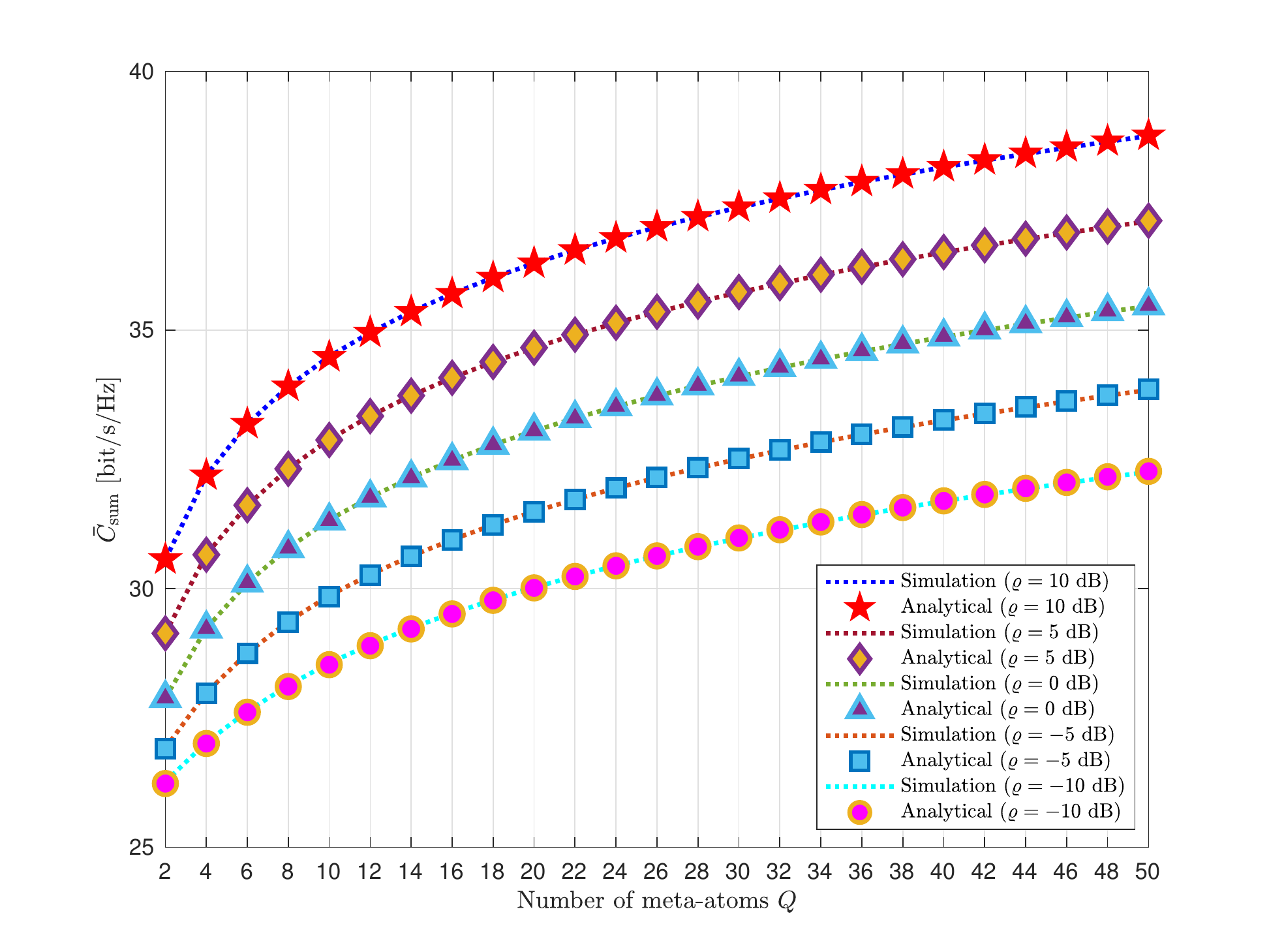}
\caption{$\overline{\capaa}_\text{sum}$ versus $Q$ ($K=10$, Approximation 2 for the analytical curves).}
\label{fig:fig_3}
\end{figure}

Fig.~\ref{fig:fig_2} and \ref{fig:fig_3} depict the performance of the RIS-aided downlink 
as a function of the number of meta-atoms $Q$ for different values of $\varrho$,
with $K=10$. In this case, $\overline{\capaa}_\text{sum}^\text{w/o RIS}=25.26$ bits/s/Hz.
In Fig.~\ref{fig:fig_2}, the rate $\overline{\capaa}_\text{sum}$ is compared with
the analytical curve 
\eqref{eq:Ave-C-sum-asympt} by using  
the parameters \eqref{eq:constant-2}
(Approximation 1), whereas 
\eqref{eq:b_K-naka} and \eqref{eq:a_K-naka} (Approximation 2)
are used in Fig.~\ref{fig:fig_3}.
Results confirm the precision of Approximation 2 and 
show that the approximation based on the hardening of the reflection
channel (Approximation 1) is inaccurate for small values of $Q$, especially when 
the reflection channel is stronger than the direct one. From the comparison 
among Figs.~\ref{fig:fig_1}, \ref{fig:fig_2}, and \ref{fig:fig_3}, it can be inferred that,
as predicted, the 
sum-rate capacity increases much faster with respect to $Q$ than $K$.


\section{Conclusions}
\label{sec:concl}

We derived two approximations of the sum-rate capacity
of an opportunistic time-sharing downlink with an RIS. The
approximation based on the hardening of the reflection 
channel allows to show the asymptotic scaling laws 
in terms of the number of users $K$ and 
the number of meta-atoms $Q$.
A more accurate approximation was derived by approximating the 
overall channel seen by each user as a gamma RV. 
Both multiuser diversity and reflection process of the RIS provide 
increased channel magnitudes. However, the SNR gain 
increases faster with respect to $Q$ than $K$ and, thus, 
the multiuser diversity effect becomes negligible even for
moderate values of $Q$. 
Herein, we have considered single-antenna
BS and receivers. An interesting research subject consists of extending
the proposed framework to the case of multi-antenna terminals.


\appendices

\section{Distributions of RVs $Z_k^{(1)}$ and $Z_k^{(2)}$}
\label{app:app-1}
The distribution of $Z_k^{(1)}$ does not depend on 
the number $Q$ of meta-atoms: it is a Rayleigh-distributed RV or, equivalently, 
it can be seen as a Nakagami RV with shape parameter $m=1$
and scale parameter $\Omega=\sigma_h^2$ \cite{Proakis}.
On the other hand, the distribution of $Z_k^{(2)}$ strongly depends on $Q$.
Indeed, it can be readily verified that $Z_k^{(2)}$ is 
a Nakagami RV with shape parameter $m=Q$
and scale parameter $\Omega=\sigma_f^2 \, \sigma_g^2 \, Q^2$, whose
mean and variance are given by (see, e.g., \cite{Proakis})
\barr
\Es[Z_k^{(2)}] & = \sigma_f \, \sigma_g \, \sqrt{Q} \, 
\frac{\Gamma\left(Q+\frac{1}{2}\right)}{\Gamma(Q)}
\approx \sigma_f \, \sigma_g  \, Q
\label{mean-Naka}
\\
\text{VAR}[Z_k^{(2)}] & = \sigma_f^2 \, \sigma_g^2 \, Q^2 
\left\{1- \frac{1}{Q} \left[
 \frac{\Gamma\left(Q+\frac{1}{2}\right)}{\Gamma(Q)} \right]^2 
\right\} 
\approx \frac{\sigma_f^2 \, \sigma_g^2 \, Q}{4} 
\label{var-Naka}
\earr
where $\Gamma(x) \eqdef \int_{0}^{+\infty} t^{x-1} \, e^{-t} \, {\rm d}t$, with $x >0$, 
is the gamma function, whereas 
the approximations come from the Stirling's series of the quotient 
$\Gamma\left(Q+\frac{1}{2}\right)/\Gamma(Q)$ \cite{Ark.1985}
for very large $Q$.

\section{Choice of the parameters $\widehat{m}$
and $\widehat{\Omega}$}
\label{app:app-2}
To obtain an accurate approximation of $f_{Z_k}(\alpha) $, 
we resort to the {\em moment matching} method. Specifically, 
the parameters $\widehat{m}$
and $\widehat{\Omega}$ are chosen such that 
to match the second and fourth moments of $Z_k$ and $\widehat{Z}_k$, i.e., 
(i) $\Es[Z_k^2]=\Es[\widehat{Z}_k^{2}]$ {\em and }
(ii) $\Es[Z_k^4]=\Es[\widehat{Z}_k^{4}]$.
By observing that $\Es[\widehat{Z}_k^{2}]=\Es[\widehat{X}_k]$,
with the mean of the gamma distribution 
\eqref{eq:Gamma-pdf} being the ratio between its shape and 
scale parameters, i.e., $2 \, \widehat{m}/(\widehat{\Omega}/\widehat{m})= 2 \, \widehat{\Omega}$,
condition (i)  yields
$\widehat{\Omega} =\frac{\Es[Z_k^2]}{2}= \frac{1}{2} \, \Es[X_k]$.
Since $\Es[\widehat{Z}_k^{4}]=\Es[\widehat{X}_k^2]$, with 
the $2$-nd  moment of the gamma RV $\widehat{X}_k$  
given \cite{Casella} by
$\Es[\widehat{X}_k^2] = \left(\frac{\widehat{\Omega}}{\widehat{m}}\right)^2 
\frac{\Gamma(2\, \widehat{m}+2)}{ \Gamma(2\, \widehat{m})} =
\frac{2 \, \widehat{\Omega}^2}{\widehat{m}} \, (2 \, \widehat{m}+1)$,
where we have also used $\Gamma(x+1)=x \, \Gamma(x)$, which is in general
valid  for all complex numbers $x$ except the non-positive integers, condition (ii) leads to
$\widehat{m} = \frac{1}{2} \, \frac{\Es^2[X_k]}{\Es[X_k^2]-\Es^2[X_k]}
=\frac{1}{2} \, \frac{\Es^2[X_k]}{\text{VAR}[X_k]}$.
By observing that 
the moments
of the Rayleigh RV $|h_k|$ and the chi-distributed RV $\|\fb_k\|$
can be expressed as (see, e.g., \cite{Proakis}) 
$\Es[|h_k|^n]  = \sigma_{h}^n \, \Gamma\left(1+\frac{n}{2}\right)$
and 
$\Es[\|\fb_k\|^n]  = \sigma_{f}^n \, \frac{\Gamma\left(Q+\frac{n}{2}\right)}{\Gamma(Q)}$,
for $n \in \mathbb{N}$,  it can be shown that the first two moments of $X_{k}$ are 
\barr
\Es[X_{k}] & = 
\sigma_{h}^2 + \sigma_{f}^2 \, \sigma_g^2 \, Q^2 + \sigma_{f} \, \sigma_g \, 
\sigma_{h} \sqrt{Q \, \pi} \, \frac{\Gamma\left(Q+\frac{1}{2}\right)}{\Gamma(Q)} 
\label{eq:meanX_k}
\\
\Es[X_{k}^2]  & = 2 \, \sigma_{h}^4 + 
3 \, \sigma_{f} \, \sigma_g \, 
\sigma_{h}^3 \, \sqrt{Q \, \pi} \, \frac{\Gamma\left(Q+\frac{1}{2}\right)}{\Gamma(Q)}
\nonumber \\ & 
+ 6 \, \sigma_{f}^2 \, \sigma_g^2 \, 
\sigma_{h}^2 \, Q^2 
+ 2 \, \sigma_{f}^3 \, \sigma_g^3 \, 
\sigma_{h} \, Q \, \sqrt{Q \, \pi} \, \frac{\Gamma\left(Q+\frac{3}{2}\right)}{\Gamma(Q)}
\nonumber \\ & 
+ \sigma_{f}^4 \, \sigma_g^4 \, Q^3 \, (Q+1)
\label{eq:rmsX_k}
\earr
where  $\Gamma(3/2)=\sqrt{\pi}/2$ and $\Gamma(5/2)=3 \, \sqrt{\pi}/4$ have been used.


\end{document}